\RequirePackage{lineno}
\newif\ifpreprint%
\preprintfalse%
\ifpreprint%
	\documentclass[preprint,english,aps,prb,floatfix,amssymb,superscriptaddress,a4paper]{revtex4-1}
\else%
	\documentclass[twocolumn,english,aps,prb,floatfix,amssymb,superscriptaddress,a4paper]{revtex4-1}
\fi%
\usepackage{amsmath}
\usepackage{physics}
\usepackage{chemformula}
\usepackage{dsfont}
\usepackage{color}
\usepackage{amssymb}
\usepackage{graphicx}
\usepackage{braket}
\usepackage{units}
\usepackage{lipsum}
\usepackage{hyperref}
\hypersetup{
  colorlinks = true,
  hypertexnames=true
}

\renewcommand{\theta}{\vartheta}
\renewcommand{\phi}{\varphi}

\begin{document}
\ifpreprint%
	\linenumbers%
\fi%

\title{Weyl orbits without an external magnetic field}

\author{
Valerio Peri}
\email{periv@phys.ethz.ch}
\affiliation{
Institute for Theoretical Physics, ETH Zurich, 8093 Z\"urich, Switzerland
}

\author{
Tena Dub\v{c}ek}
\affiliation{
Institute for Theoretical Physics, ETH Zurich, 8093 Z\"urich, Switzerland
}

\author{
Agnes Valenti}
\affiliation{
Institute for Theoretical Physics, ETH Zurich, 8093 Z\"urich, Switzerland
}

\author{
Roni Ilan}
\affiliation{
Raymond and Beverly Sackler School of Physics and Astronomy, Tel-Aviv University, Tel-Aviv 69978, Israel
}

\author{
Sebastian D. Huber}
\affiliation{
Institute for Theoretical Physics, ETH Zurich, 8093 Z\"urich, Switzerland
}

\begin{abstract}
Weyl semimetals in a magnetic field give rise to interesting non-local electronic orbits: the ballistic transport through the bulk enabled by the chiral Landau levels is combined with a momentum-space sliding along the surface Fermi-arc driven by the Lorentz force. Bulk chiral Landau levels can also be induced by axial fields whose sign depends on the chirality of the Weyl point. However, the microscopic perturbations that give rise to them can be described in terms of gauge fields only in the low-energy sectors around the Weyl points. In addition, since pseudo-fields are intrinsic, there is no apparent reason for a Lorentz force that causes sliding along the Fermi-arcs. Therefore, the existence of non-local orbits driven exclusively by pseudo-fields is not obvious. Here, we show that for systems with at least four Weyl points in the bulk spectrum, non-local orbits can be induced by axial fields alone. We discuss the underlying mechanisms by a combination of analytical semi-classical theory, the microscopic numerical study of wave-packet dynamics, and a surface Green's function analysis.
\end{abstract}

\date{\today}

\maketitle
\section{Introduction}
\label{sec:intro}

Controlling the flow of electrons by taking advantage of the band structure topology is an interesting endeavor both from a fundamental and a practical perspective \cite{Hasan:2010,Qi:2011}. In two-dimensional topological materials, electrons are guided through unidirectional scattering-free channels confined at the system surface. Unlike the integer quantum Hall effect, where time-reversal symmetry is broken by external magnetic fields \cite{Thouless:1982}, the surface channels of topological insulators appear also in the presence of time-reversal symmetry and do not require external fields \cite{Konig07,Bernevig06a,Kane:2005}. 

In three dimensions, an alternative avenue to engineer the electronic flow is offered by Weyl semimetals \cite{Armitage:2018, Burkov:2016, Burkov:2011, Lv:2015, Xu:2015a, Yang:2015, Lu:2015}. These systems lack a full gap in their bulk spectrum. Instead, two energy bands touch at isolated points in reciprocal space. The quasiparticle excitations around these nodal points have a linear dispersion and resemble massless relativistic Weyl fermions. What makes these systems particularly interesting is a topological charge associated with the spectral degeneracies: each Weyl point acts as a source or sink of Berry curvature according to its chirality. The total Berry curvature in a 3D Brillouin zone must vanish. Therefore, Weyl points always come in pairs of opposite chirality in any lattice system \cite{Nielsen:1983}.

As much as for the integer quantum Hall effect and topological insulators, the non-trivial bulk topology manifests itself at the sample surfaces \cite{Armitage:2018,Burkov:2016,Burkov:2011}. Namely, open equi-energy contours connect the projections of Weyl points of opposite chirality on the surface Brillouin zone and realize unidirectional channels on the surface of Weyl semimetals: the Fermi arcs.

In addition to the non-trivial surface physics, Weyl semimetals react in intriguing ways to the application of external fields. A magnetic field discretizes the spectrum in relativistic Landau levels and a unique feature of Weyl semimetals is the presence of a zeroth Landau level that disperses along the field direction and has zero group velocity perpendicular to it \cite{Nielsen:1983}. In particular, the sign of the dispersion of the chiral channel depends on the chirality of the unperturbed Weyl point. Hence, bulk unidirectional channels are separated in momentum space rather than real space, making Weyl systems in a magnetic field the 3D reciprocal space counterpart of the real space 2D integer quantum Hall effect. 

The simultaneous presence of open Fermi-arcs ending at the Weyl-point projections on the surface and the chiral bulk channels is the key ingredient of a unique magneto-transport signature of Weyl systems: Weyl orbits \cite{Potter:2014,Baum:2015,Zhang:2016,Borchmann:2017,Wang:2017,Yao:2017,McCormick:2018,Pareek:2018,Pikulin:2020}. They are non-local, closed trajectories that mix bulk and boundary degrees of freedom. The electrons with momentum around a Weyl point propagate ballistically through the bulk along (or against) the field direction on the field-induced chiral channel. Once they reach the surface, they slide along the Fermi arc under the action of the Lorentz force until they reach the projection of a Weyl point of opposite chirality. There, the chiral zeroth Landau level propagating in the opposite direction carries them through the bulk until they reach the opposite surface and the process repeats itself. This motion gives rise to a conveyor belt-like motion that leads to a transmission of electrons through a Weyl semimetal \cite{Baum:2015}, cf. Fig.~\ref{fig:fig2}(a).

Recently, experimental evidences have been put forward for Weyl orbits in Dirac semimetals \cite{Moll:2016, Zhang:2017, Uchida:2017, Zhang:2018}. These materials have four-fold degenerate points in their spectrum which can be seen as two Weyl points of opposite chirality superimposed in momentum space \cite{Armitage:2018}. The lack of a well-defined chirality associated with Dirac points complicates the interpretation of these experimental results in terms of Weyl orbits.

Magnetic fields are not the only example of gauge fields that couple to Weyl fermions. An intriguing feature of the Weyl semimetals is that the two different chiral flavors have an independent gauge degree of freedom. In principle, it is possible to envision an axial gauge field that couples with different signs to Weyl points of opposite chirality. Such a field is tightly connected to the axial anomaly studied in high-energy physics \cite{Landsteiner:2016, Bell1969, Adler:1969}, but no axial background field is present in quantum electrodynamics. On the other hand, it has been argued that a similar phenomenology can be achieved in condensed matter settings via, for example, inhomogeneous uniaxial strain \cite{Ilan:2019,Zhou:2013, Liu:2013, Chernodub:2014, Cortijo:2015, Grushin:2016, Sumiyoshi:2016, PhysRevB.94.241405, Pikulin:2016, Liu:2017, Gorbar:2017, Alisultanov:2018, Roy:2018, Behrends:2019, Kamboj:2019, Gorbar:2017a, Behrends:2019a}. It is important to stress that all the proposed implementations mimic gauge fields exclusively in the low-energy sector and the ``gauge'' choice leads to observable effects. This is why the axial field in condensed matter systems is also dubbed the pseudo-magnetic field. Recently, the axial field has been realized in carefully engineered photonic \cite{Jia:2019} and acoustic \cite{Peri:2019} structures that simulate the phenomenology of electronic Weyl semimetals. 

The chirality-dependent coupling of an axial field has important consequences for the Landau levels structure. Namely, in a two Weyl points system it leads to co-propagating chiral channels rather than counter-propagating ones \cite{Grushin:2016}. What is the fate of Weyl orbits in the presence of an axial field? In the following, we seek the answer to this question. Clearly, they cannot take place in a minimal two Weyl points model. The co-propagating zeroth Landau levels render the bulk propagation fully chiral, meaning, no way back through the bulk is available. Instead, the electrons remain confined at the surface. This situation has been argued to give rise to a topological coaxial cable \cite{Pikulin:2016}. The scenario drastically changes in the presence of time-reversal symmetry. In this case, the minimal number of Weyl points is four \cite{Nielsen:1983}. Moreover, the co-propagating bulk channels of a pair of Weyl points are accompanied by the counter-propagating set of zeroth Landau levels associated with the time-reversal partner pairs. In fact, time-reversal symmetry does not allow for fully chiral bulk channels. As we show in this work, Weyl orbits driven exclusively by an axial field are in principle possible in the time-reversal symmetric setting, albeit not obvious. This, despite the fact that the pseudo-magnetic field cannot be described in terms of gauge fields away from the low-energy sector of the Weyl points and an interpretation of its effect in terms of a Lorentz force is at best in question. Moreover, the connectivity of the Fermi arcs might prevent closed bulk-boundary oscillations, as we will further argue in this work. 

The remainder of the paper is organized as follows. In Sec.~\ref{sec:model}, we introduce the minimal model for a time-reversal symmetric Weyl system used in this work. In Sec.~\ref{sec:axial}, we discuss the difference between magnetic and axial fields. We introduce a pseudo-magnetic field in our lattice model and study its effects on the bulk physics. In Sec.~\ref{sec:arcs}, we carefully characterize the surface physics of the model via effective surface Green's functions and show how the arc connectivity can be tuned via a simple parameter \cite{Dwivedi:2016}.
Compelling evidence for axial-field-induced Weyl orbits is provided in Sec.~\ref{sec:wp}~and~\ref{sec:dos}. First, we perform numerical simulations of the exact dynamical evolution of wave-packets in a finite lattice system. Second, we study the Bohr-Sommerfeld quantization induced by the closed Weyl orbits via effective surface Green's functions. Some concluding remarks are given in Sec.~\ref{sec:conclude}.

\section{Minimal time-reversal Weyl system}
\label{sec:model}
We focus our attention on a Weyl system, namely a system with Weyl points in its spectrum, that preserves time-reversal symmetry and breaks inversion symmetry. Such a system could be an electronic material, but also a classical metamaterial or an ultracold atoms setup. As prescribed by the Nielsen-Ninomiya theorem \cite{Nielsen:1983}, one will find at least four Weyl points in the Brillouin zone. 

Our results are generic and apply to any time-reversal symmetric Weyl systems with an arbitrary number of Weyl nodes. To illustrate concretely and quantitatively the key ideas of our work, however, we introduce a minimal model that was originally put forward in Ref.~[\onlinecite{Dwivedi:2016}]. Such a model has various advantages from a theoretical standpoint. Namely, as we will show below, it is possible to tune the connectivity of the Fermi arcs on the surface via a single parameter. Moreover, two other parameters allow to tune the location of the four Weyl points in momentum space.

The two-band Bloch Hamiltonian describing the tight-binding model of Ref.~[\onlinecite{Dwivedi:2016}] is:
\begin{equation}
	\label{eqn:model}
H(\mathbf{k})=H_x(k_x)+H_\perp(\mathbf{k}_\perp)\,,	
\end{equation}
where
\begin{align}
	H_x(k_x)&=\tau\sin{k_x}\sigma^x-\tau\cos{k_x}\sigma^z\,,\\
	\label{eqn:Hperp}
	H_\perp(\mathbf{k}_\perp)&=\gamma\left({\mathbf{k}_\perp}\right)\mathbb{I}+\tau\eta_y(\mathbf{k}_\perp)\sigma^y+\tau\left[ 1+ \eta_z(\mathbf{k}_\perp)\right]\sigma^z\,,
\end{align}
and $\gamma\,,\eta_y\,,\eta_z\,:\mathbb{T}^2\rightarrow \mathbb{R}$ are functions of the ``transverse'' momentum $\mathbf{k}_\perp=(k_y\,,k_z)$, $\tau$ is the hopping parameter and the Pauli matrices $\sigma^i$ act on an orbital degree of freedom.

Weyl nodes appear if one considers:
\begin{equation}
	\begin{pmatrix}
	\eta_y \\
	\eta_z
	\end{pmatrix}=
	M\begin{pmatrix}
	\cos{k_y}-\cos{b_y}\\
	\cos{k_z}-\cos{b_z}
	\end{pmatrix}\,,
\end{equation}
where $M\in \text{SL}(2,\mathbb{R})$ and $b_z,\,b_y\in[0,\pi]$. The location of the Weyl points is independent of the choice of $M$ and entirely determined by $b_z$ and $b_y$. For $b_z\neq 0$ and $b_y\neq 0$, the four Weyl nodes are found at $(k_x^*\,,k_y^*\,,k_z^*)=(0\,,\pm b_y\,,\pm b_z)$. There, $\eta_y(k_y^*\,,k_z^*)=0$, $\eta_z(k_y^*\,,k_z^*)=0$ and $H_x(k_x^*)+\tau\sigma^z=0$. The chirality $\chi$ is given by $\chi=
{\rm sign}[-\tau\sin{k_y^*} \sin{k_z^*}]$.

The details of the matrix $M$ leave the bulk spectrum invariant but affect the curvature and connectivity of the Fermi arcs, as further discussed in Sec.~\ref{sec:arcs}. In this work, following Ref.~[\onlinecite{Dwivedi:2016}], we choose 
\begin{equation}
	\label{eqn:Mmatrix}
	M=\begin{pmatrix}
	\cos{\varphi} & -\sin{\varphi}\\
	\sin{\varphi} & \phantom{-}\cos{\varphi}
	\end{pmatrix}\,,
\end{equation}
with $\varphi\in[\pi,3/2\pi]$. 

Whenever the function $\gamma$ has a non-trivial dependence on ${\mathbf{k}_\perp}$, the Weyl cones get tilted. Here, we assume $\gamma\left({\mathbf{k}_\perp}\right)=\epsilon_0$ and only a global energy shift $\epsilon_0$ is allowed. Namely, we deal with ideal type-I Weyl semimetals with non-tilted Weyl points all at the same energy \cite{Armitage:2018}.

Let us stress once again that the choice of this particular model is driven exclusively by its theoretical appeal, as it allows us to highlight our main results in a concrete example. It does not attempt to describe any known material nor is it particularly suitable for implementation in engineered platforms. Nevertheless, none of the results presented in this work crucially hinges on the peculiarities of this model. Rather, our findings are generic and can be reformulated for arbitrary time-reversal symmetric tight-binding models with closer connections to electronic Weyl semimetals or Weyl systems in engineered platforms.

\section{Inhomogeneous Weyl semimetals and axial field}
\label{sec:axial}
\subsection{Weyl semimetals and gauge fields: magnetic vs. axial}
\label{ssec:twoaxial}
\begin{figure}[t]
\includegraphics[]{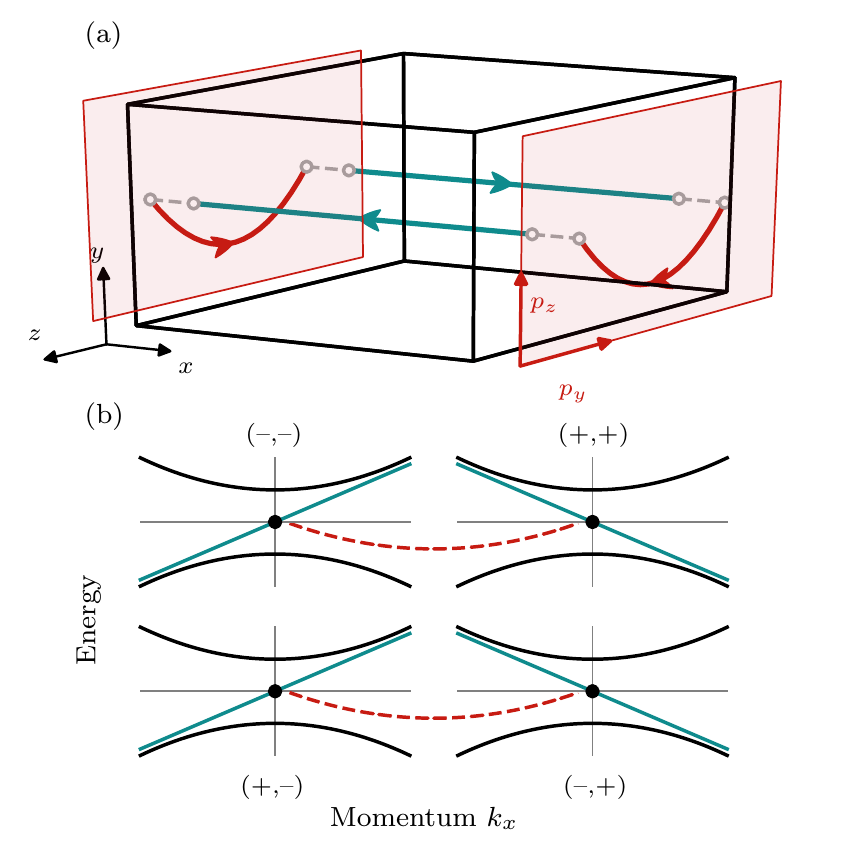}
\caption{\label{fig:fig2} (a) Schematic of a Weyl orbit. Green lines represent real space motion through the bulk, while red ones show the sliding on the Fermi arcs in the surface Brillouin zone. (b) Chiral channels induced in the model of Eq.~\eqref{eqn:model} by the axial field of Sec.~\ref{ssec:axial4}. The red dashed lines indicate the Fermi arc connectivity that allows for Weyl orbits. The tuple $(\chi,\xi)$ (see main text) entirely characterizes each of the four Weyl points in the system and it is shown for each chiral channel.}
\end{figure}
Some of the most intriguing aspects of Weyl semimetals are their transport properties in the presence of gauge fields. In this Section, we will review how the coupling to these fields affects the bulk physics. 

A low-energy description of an optimally doped Weyl semimetal is particularly suitable to study these properties. Indeed, around the Weyl points energy $\epsilon_0$, the bulk properties of a Weyl system with two Weyl points are dominated by the quasi-particles described by a low-energy Weyl Hamiltonian: 
\begin{equation}
	\label{eqn:hlowe}
	H_{2,\text{low}}(\mathbf{k})=\sum_{i,j}\chi v_{ij}\left(k_i-\chi A_{5i}\right)\sigma^j\,,
\end{equation}
where $v_{ij}$ denotes the velocity tensor, the Pauli matrices $\sigma^i$ encode a pseudo-spin degree of freedom, the vector $2\mathbf{A}_5$ indicates the Weyl points separation in reciprocal space and $\chi=\pm$ is the chirality of the Weyl points. Eq.~\eqref{eqn:hlowe} provides a low-energy description of the minimal model of Weyl fermions. Namely, the  Nielsen-Ninomiya theorem \cite{Nielsen:1983} requires a minimum of two Weyl points of opposite chirality, which are here separated by $2\mathbf{A}_5$ in reciprocal space.

An external magnetic field $\mathbf{B}$ can be directly added via minimal coupling: $\mathbf{k}\rightarrow \mathbf{k}-e\mathbf{A}$, where $\mathbf{A}$ is the electromagnetic gauge potential and $\mathbf{B}=\nabla\times \mathbf{A}$. For simplicity, we consider the concrete case $\mathbf{B}=B\hat{\mathbf{x}}$ and choose to work in the Landau gauge $\mathbf{A}=By\hat{\mathbf{z}}$. 

The external field alters the energy spectrum giving rise to discrete Landau levels \cite{Nielsen:1983}:
\begin{equation}
	\label{eqn:specField}
	\epsilon_n=\begin{cases}
		 \chi\text{sign}(B)v_Fk_x & n=0 \,,\\
		 \pm v_F\sqrt{k_x^2+2|n|eB} & n\neq0 \,,
		\end{cases}
	\end{equation}
with $n\in\mathbb{Z}$. For $n\neq 0$, relativistic Landau levels with energy spacing $\propto \sqrt{n}$ appear. The unique feature of Weyl semimetals is the appearance of chiral zeroth Landau levels linearly dispersing along (or against) the field. In particular, Weyl points of opposite chirality $\chi$ have counter-propagating zeroth Landau levels. 

From a semiclassical perspective, an external magnetic field shifts the Weyl points in momentum space as a function of real space: $\mathbf{k}\rightarrow \mathbf{k}-e\mathbf{A}$. The vector $\mathbf{A}_5$ enters the low-energy description of Eq.~\eqref{eqn:hlowe} similarly to a gauge potential: $\mathbf{k}\rightarrow \mathbf{k}-\chi\mathbf{A}_5$. Whenever $\mathbf{A_5}$ becomes spatially dependent, also the location of Weyl nodes in momentum space becomes a function of real space. In comparison to a real magnetic field, however, nodes of opposite chirality shift in opposite directions. A space-dependent $\mathbf{A}_5$ induces an axial or pseudo-magnetic field: $\mathbf{B}_5=\nabla\times\mathbf{A}_5$. This axial field can be regarded as a magnetic field that couples with opposite signs to nodes of opposite chirality.

Similarly to an external magnetic field, one could study the transport properties of Weyl semimetals in the presence of an axial field. There is, however, a key difference: the gauge potential $\mathbf{A}_5$ couples to the topological charge $\chi$ rather than the electric charge $e$. This alters the discretization of the spectrum in Landau levels \cite{Grushin:2016}:
\begin{equation}
	\label{eqn:spec5Field}
	\epsilon_n=\begin{cases}
		 \text{sign}(B_5)v_Fk_x & n=0 \,,\\
		 \pm v_F\sqrt{k_x^2+2|n|eB_5} & n\neq0 \,,
		\end{cases}
	\end{equation}
with $n\in\mathbb{Z}$. Note that the dispersion of the zeroth chiral Landau level ceases to depend on the chirality $\chi$. Weyl points of opposite chirality give rise to co-propagating chiral channels. This seemingly innocent change has profound effects on many transport phenomena, notably on Weyl orbits. 

A question of paramount importance is whether and how it is possible to realize this pseudo-magnetic field. This amounts to a spatially dependent separation of the Weyl points in reciprocal space $\mathbf{A}_5=\mathbf{A}_5(\mathbf{r})$, such that $\nabla\times\mathbf{A}_5\neq 0$. Various proposals have been put forward. For example, inhomogeneous strain \cite{Grushin:2016,Pikulin:2016,Liu:2017,PhysRevLett.116.166601,Cortijo:2015,Kamboj:2019,Destraz:2020}, inhomogeneous magnetization \cite{PhysRevB.87.235306,PhysRevB.94.115312} or propagating sound waves \cite{PhysRevB.94.241405,Sukhachov:2020} (see Ref.~[\onlinecite{Ilan:2019}] for a review). The axial field does not couple to the electromagnetic charge, rather to the topological charge $\chi$. As such, it is amenable to realization also in platforms with bosonic neutral excitations as neutral ultracold atoms and phononic or photonic crystals. 

The concrete implementation of an axial field points to the crucial difference from an external magnetic field. The latter is an external gauge field that minimally couples to momentum across all energy scales. The former is a complicated microscopic perturbation that acts as a gauge field exclusively in the low-energy description around a Weyl point as in Eq.~\eqref{eqn:hlowe}. In turn, this means that the choice of ``gauge'' can have observable consequences. 

\subsection{Axial field in a time-reversal four Weyl points system}\label{ssec:axial4} 
To study whether Weyl orbits can occur in the exclusive presence of an axial field, we introduce it in the model of Eq.~\eqref{eqn:model}. At least at a theoretical level, this can be readily done. 

By promoting $b_y$ and $b_z$ to space-dependent functions, the location of the Weyl nodes in momentum space acquires a spatial dependence. In particular, Weyl points of opposite chirality shift in opposite directions: the minimal requirements for the low-energy physics to be captured by an axial gauge field. 

We consider a field along $\hat{\mathbf{x}}$ and choose to work in the Landau gauge with an axial gauge potential $\mathbf{A}_5=B_5y\hat{\mathbf{z}}$. Therefore, we add to the microscopic theory the following space-dependent perturbations:
\small
\begin{equation}
	\label{eqn:axial1}
	\begin{split}
\eta_y=&\cos(\varphi)\left[\cos(k_y)-\cos(b_y^0)\right]\\&-\sin(\varphi)\left[\cos(k_z)-\cos\left(b_z^0-B_5\left(y-\frac{L_y}{2}\right)\right)\right]\,,
\end{split}
\end{equation}
\normalsize
and
\small
\begin{equation}
	\label{eqn:axial2}
	\begin{split}
\eta_z=&\sin(\varphi)\left[\cos(k_y)-\cos(b_y^0)\right]\\&+\cos(\varphi)\left[\cos(k_z)-\cos\left(b_z^0-B_5\left(y-\frac{L_y}{2}\right)\right)\right]\,.
\end{split}
\end{equation}
\normalsize
This change in the microscopic formulation of Eq.~\eqref{eqn:model} highlights how the perturbation acts as a gauge field exclusively in the low-energy theory. At the same time, it captures the essential ingredient of an axial field without specifying its physical origin. This feature makes the chosen tight-binding model particularly suitable to keep the discussion as generic as possible. Our results hold in the presence of an axial field of arbitrary origin, regardless of whether it was induced by inhomogeneous strain or by other means. 

The Weyl points are moved around their original unperturbed positions given by $(b_y^0\,,b_z^0)=(\pi/2\,,\pi/2)$. This ensures that the chiral Landau levels induced by the axial gauge field are centered at $(0\,,\pm b_y^0\,,\pm b_z^0)$.
In particular, the channels velocity is towards $+\hat{\mathbf{x}}$ at $(0\,,-b_y^0\,,\pm b_z^0)$, while towards $-\hat{\mathbf{x}}$ at $(0\,,+b_y^0\,,\pm b_z^0)$, cf. Fig.~\ref{fig:fig2}(b). The presence of time-reversal symmetry forbids fully chiral bulk-channels: time-reversal $\mathcal{T}$ acts on the Bloch Hamiltonian as $\mathcal{T}H^*(\mathbf{k})\mathcal{T}=H(-\mathbf{k})$, thereby, it reverses the dispersion of the unidirectional chiral channels described by the effective Hamiltonian: $H(k_x)=v_Fk_x$. This situation is more complicated than the inversion symmetry preserving setting considered in Sec.~\ref{ssec:twoaxial} and Eq.~\eqref{eqn:spec5Field} and allows for new intriguing phenomena. 

In the presence of four Weyl points, it is useful to introduce an additional $\mathbb{Z}_2$ index $\xi$ to distinguish among time-reversal partners. For the model of Eq.~\eqref{eqn:model}, we choose $\xi=+1$ ($\xi=-1$) for Weyl points at $+b^0_y$ ($-b^0_y$). The minimal coupling of an axial gauge field for a time-reversal symmetric system with four Weyl points is captured by $\mathbf{k}\rightarrow\mathbf{k}-\xi\chi\mathbf{A}_5$. This can be appreciated in the low-energy expansion of Eq.~\eqref{eqn:model} in the presence of the space-dependent perturbations of Eq.~\eqref{eqn:axial1}--\eqref{eqn:axial2}: 
\small
 \begin{align}
 	 \label{eqn:lowOur}
	\begin{split}
		H_\text{low}(\mathbf{k})=&+\tau \left[\beta\xi\chi\sin{\phi} \left(k_z-\xi\chi B_5\left(y-\frac{L_y}{2}\right)\right)\right.\\&\left. \vphantom{\beta\xi\chi\sin{\phi} \left(k_z-\xi\chi B_5\left(y-\frac{L_y}{2}\right)\right)}\qquad -\xi\cos{\phi}k_y \right]\sigma^y \\& -\tau\left[\beta\xi\chi\cos{\phi} \left(k_z-\xi\chi B_5\left(y-\frac{L_y}{2}\right)\right)\right. \\&  \left. \vphantom{\beta\xi\chi\cos{\phi} \left(k_z-\xi\chi B_5\left(y-\frac{L_y}{2}\right)\right)}\qquad  +\xi\sin{\phi}k_y\right]\sigma^z \\&  +\tau k_x \sigma^x\,,
	\end{split}
  \end{align}
\normalsize
where we chose $b_y^0=b_z^0=\pi/2$. Here, $\xi$ and $\chi$ are the $\mathbb{Z}_2$ indices introduced to describe a minimal Weyl system in the presence of time-reversal symmetry, while $\beta=\cos{\left[B_5\left(y-L_y/2\right)\right]}$ is a space-dependent renormalization of the Fermi velocity caused by the axial potential. This velocity correction stresses how the microscopic origin of the axial pseudo-potential can lead to observables effects of the ``gauge'' choice.  


\section{Surface physics \label{sec:arcs}}
\subsection{Effective surface Green's functions\label{ssec:eff}}
Weyl orbits rely on the presence of open equi-energy contours on the sample surface that connect the projection of Weyl points of opposite chirality. In this Section, we carefully characterize the surface physics of the model of Eq.~\eqref{eqn:model}.

A variety of methods has been recently put forward to obtain the surface states of Weyl semimetals \cite{Borchmann:2017,Dwivedi:2016,Pinon:2020,Kaladzhyan:2020}. Here, we start by going back to a situation without an applied (axial) field and rely on effective surface Green's functions to study the Fermi arcs of our model \cite{Borchmann:2017,Marchand:2012}.
The Weyl nodes are separated in the $k_y$--$k_z$ plane. Therefore, we expect Fermi arcs connecting the projections of Weyl points of opposite chirality $\chi$ on surfaces perpendicular to $\hat{\mathbf{x}}$. 

We consider a slab finite in the $\hat{\mathbf{x}}$ direction and infinite along $\hat{\mathbf{y}}$ and $\hat{\mathbf{z}}$, such that $k_y$ and $k_z$ continue to be good quantum numbers. Each of the $L_x$ layers along $\hat{\mathbf{x}}$ is described by the $2\times 2$ Bloch Hamiltonian $H_\perp(\mathbf{k}_\perp)$ of Eq.~\eqref{eqn:Hperp}. Each layer is then coupled to the neighboring ones via the matrix
\begin{equation}
	\label{eqn:rmatrix}
	R=\begin{pmatrix}
	-\phantom{i}\tau/2 & -i\tau/2\\
	-i\tau/2 & \phantom{-}\phantom{i}\tau/2 
	\end{pmatrix}\,,
\end{equation}
and its conjugate transpose $R^\dagger$.

The Hamiltonian describing the whole slab is a $2L_x \times 2L_x$ matrix:
\small
\begin{equation}
	\label{eqn:slab}
 	H_\text{slab}(\mathbf{k}_\perp)=\left(
	\begin{smallmatrix}
	 H_\perp(\mathbf{k}_\perp) & R & 0 & 0 & \dots & 0 & 0\\
	 R^\dagger & H_\perp(\mathbf{k}_\perp) & R & 0 & \dots & 0 & 0\\
	 0 & R^\dagger & H_\perp(\mathbf{k}_\perp) & R & \dots & 0 & 0\\
	 \vdots & \vdots & \vdots & \vdots & \ddots & \vdots & \vdots\\
	 0 & 0 & 0 & 0 & \dots & R^\dagger & H_\perp(\mathbf{k}_\perp)\\
	\end{smallmatrix}\right)\,.
\end{equation}
\normalsize
Rather than considering the entire system, we can attempt a separate description of the surface and bulk physics. The former consists only of the upper and lower layers. In principle, it is captured by the Hamiltonian:
\begin{equation}
	H_s(\mathbf{k}_\perp)=\begin{pmatrix}
	H_\perp(\mathbf{k}_\perp)&0\\
	0&H_\perp(\mathbf{k}_\perp)\\
	\end{pmatrix}\,,
\end{equation}
where the upper (lower) block describes the upper (lower) surface. The bulk, on the other hand, is represented by a matrix $H_b(\mathbf{k}_\perp)$ analogous to Eq.~\eqref{eqn:slab} but of size $2(L_x-2)\times 2(L_x-2)$, i.e. the whole slab with the external layers removed. From the surface and bulk Hamiltonians, we can define the respective decoupled Green's function in the Matsubara representation:
\begin{equation}
	\label{eqn:bareGreen}
	G_{b,s}(i\omega_n,\mathbf{k}_\perp)=-\left[i\omega_n-H_{b,s}(\mathbf{k}_\perp)\right]^{-1}\,,
\end{equation}
where $\omega_n=(2n+1)\pi/\beta$ with $n \in \mathbb{N}$ and $\beta=1/k_BT$. 

The bulk and the surfaces are coupled via the $2(L_x-2)\times 4$ matrix:
\begin{equation}
	T=\begin{pmatrix}
	R^\dagger & 0 \\
	0 & 0 \\
	\vdots & \vdots \\
	0 & R\\
	\end{pmatrix}\,.
\end{equation}
Therefore, an effective description of the surface physics requires to carefully trace out the bulk and not to simply neglect it. Especially in the vicinity of the Weyl points, bulk and surface modes are highly coupled and a description in terms of the simple bare surface Green's function of Eq.~\eqref{eqn:bareGreen} is not sufficient. This leads to an effective surface Green's function \cite{Borchmann:2017,Marchand:2012}:
\small
\begin{equation}
	\label{eq:ssss}
	G_{\text{eff}}(i\omega_n,\mathbf{k}_\perp)=[G_s^{-1}(i\omega_n,\mathbf{k}_\perp)-T^\dagger G_b(i\omega_n,\mathbf{k}_\perp)T]^{-1}\,.
 \end{equation}
 \normalsize
Note that this effective surface propagator contains a finite lifetime induced by the possible decay of surface states into the bulk and cannot be directly related to an effective surface Hamiltonian. From the effective surface Green's function, the surface spectral density can be directly computed as $S_\text{eff}(i\omega_n,\mathbf{k}_\perp)=-\frac{1}{\pi}\text{Im}\left[\text{Tr}\left(G_{\text{eff}}(i\omega_n,\mathbf{k}_\perp)\right)\right]$.
Eq.~\eqref{eq:ssss} is the simple equation that allows us to exactly locate the Fermi arcs of our model as explained in the following.

\subsection{Fermi arcs}
\label{ssec:arcsExact}
Via the effective surface Green's function of Sec.~\ref{ssec:eff}, we can readily show how the connectivity of the Fermi arcs can be tuned through the parameter $\varphi$ of Eq.~\eqref{eqn:Mmatrix} \cite{Dwivedi:2016}. This will prove to be a crucial factor for the existence of Weyl orbits induced by an axial field. 

Fermi arcs are the surface manifestation of the non-trivial bulk topology of Weyl semimetals. They can be understood by reducing a Weyl semimetal to a set of Chern insulators. Concretely, transversing a conical touching point in momentum space changes the Chern number of the two-dimensional momentum-space slices perpendicular to the direction in which the Weyl point is crossed. In other words, a layer Chern number acquires a non-zero value in between the projection of two Weyl points of opposite chirality. Consequently, the number of chiral edge channels per momentum space layer changes. This abrupt change in the number of surface states is manifested by the open Fermi arcs. Alternative interpretations of Fermi arcs that do not rely on a layer Chern number have been recently discussed \cite{Grushin:2016,Tchoumakov:2017,Peri:2018,Ilan:2019}.

Generically, one has to resort to numerical methods to obtain the effective surface Green's function of a finite slab and locate the Fermi arcs as described in Sec.~\ref{ssec:eff}. Nonetheless, the study of a semi-infinite sample allows for an analytical approach. In this case, we can exactly locate the Fermi arcs as a function of the parameter $\varphi$ of Eq.~\eqref{eqn:Mmatrix}.

\begin{figure}[t]
\includegraphics[]{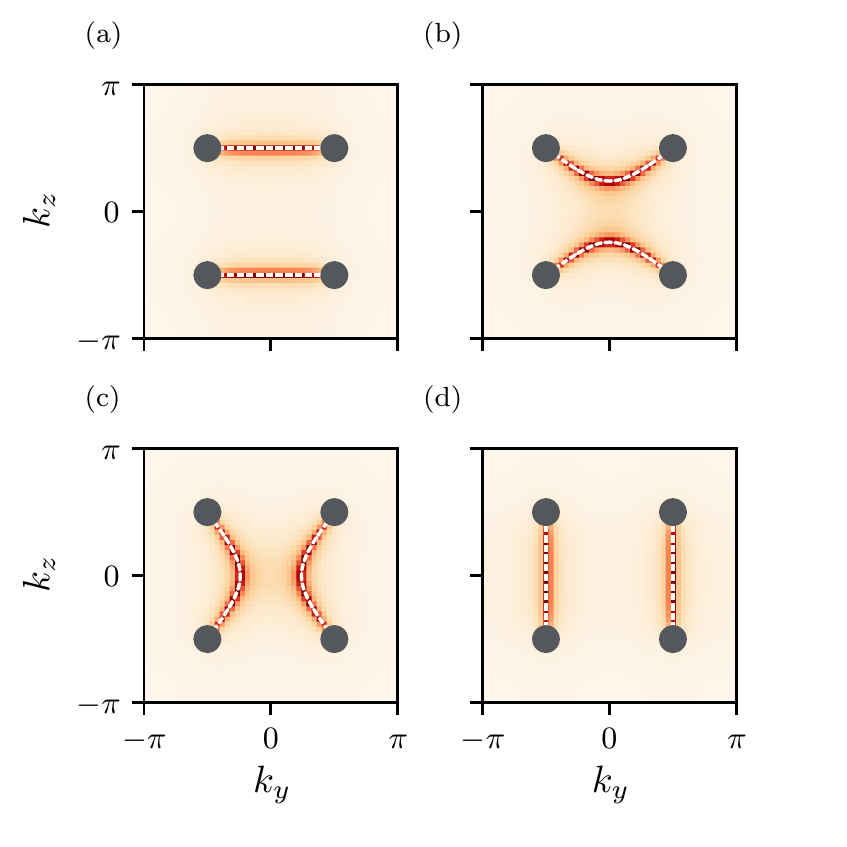}
\caption{\label{fig:fig1} Surface Brillouin zone with Fermi arcs for the model of Eq.~\eqref{eqn:model}. The grey dots represent the location of the Weyl points for $b_y=b_z=\pi/2$. The dashed lines show the analytical prediction of Eq.~\eqref{eqn:arcs}, while the colormap is the effective surface spectral density obtained from a finite slab. Different panels present results for different values of $\varphi$: in (a) $\varphi=1.5\pi$, in (b) $\varphi=1.3\pi$, in (c) $\varphi=1.2\pi$, in (d) $\varphi=\pi$. }
\end{figure}

We consider the limit where $L_x\rightarrow\infty$ and $H_s(\mathbf{k}_\perp)=H_\perp(\mathbf{k}_\perp)$. Removing one layer, e.g., the surface layer, in a semi-infinite slab does not change the sample. Therefore, we can write a self-consistent equation for the effective surface Green's function:
\small
\begin{equation}
	\label{eqn:semiInf}
	G_{\text{eff}}(i\omega_n,\mathbf{k}_\perp)=[G_\perp^{-1}(i\omega_n,\mathbf{k}_\perp)-R^\dagger G_{\text{eff}}(i\omega_n,\mathbf{k}_\perp)R]^{-1}\,,
\end{equation}
\normalsize
where $G_\perp(i\omega_n,\mathbf{k}_\perp)=-\left[i\omega_n-H_\perp(\mathbf{k}_\perp)\right]^{-1}$.

The calculations simplify performing a unitary rotation $U=\text{exp}\left(-i\pi \sigma^x/4\right)$ such that:
\begin{equation}
	U^\dagger RU=\begin{pmatrix}
	\phantom{-}0 & 0\\
	-i\tau & 0\\
	\end{pmatrix}\,,
\end{equation}
and
\begin{equation}
	U^\dagger H_\perp(\mathbf{k}_\perp)U=\epsilon_0\mathbb{I}-\tau\eta_y(\mathbf{k}_\perp)\sigma^z+\tau\tilde{\eta}_z(\mathbf{k}_\perp)\sigma^y\,,
\end{equation}
where $\tilde{\eta}_z(\mathbf{k}_\perp)=1+\eta_z(\mathbf{k}_\perp)$.

In the following analysis, we assume $\epsilon_0=0$ and set the arbitrary energy scale $\tau=1$. We solve Eq.~\eqref{eqn:semiInf} to obtain the different components of $G_\text{eff}$:
\small
\begin{align}
	G_{\text{eff}}^{(1,1)}&=\frac{\left(i\omega_n-\eta_y\right)\left[-(i\omega_n)^2+(1+\eta_y^2+\tilde{\eta}_z^2)\pm\sqrt{p}\right]}{2\tilde{\eta}_z^2}\,,\\
	G_{\text{eff}}^{(1,2)}&=i\frac{(i\omega_n)^2-(1+\eta_y^2+\tilde{\eta}_z^2)\mp \sqrt{p}}{2\tilde{\eta}_z}\,,\\
	G_{\text{eff}}^{(2,2)}&=\frac{-(i\omega_n)^2-(1-\eta_y^2-\tilde{\eta}_z^2)\pm\sqrt{p}}{2\left(i\omega_n-\eta_y\right)}\,,
\end{align}
\normalsize
where
\begin{equation}
	p=4\left(-(i\omega_n)^2+\eta_y^2\right)+\left[-(i\omega_n)^2+(\eta_y^2+\tilde{\eta}_z^2-1)\right]^2\,.
\end{equation}
Analytic continuation $i\omega_n\rightarrow \omega+i\delta$ gives the retarded Green's function $G_\text{eff}^R$. In the low-energy physics, the dominant contribution comes from the poles of $G_\text{eff}^{(2,2)R}$ at $\omega\rightarrow \eta_y$. We thus obtain:
\begin{equation}
	G_\text{eff}^{(2,2)R}=\frac{\tilde{\eta}_z^2-1-\lvert\tilde{\eta}_z^2-1\rvert}{2(\omega+i\delta-\eta_y)}\,,
\end{equation}
The spectral function is:
\small
\begin{equation}
	S_\text{eff}(\omega,\mathbf{k}_\perp)\propto\begin{cases}
	 \left[\tilde{\eta}_z(\mathbf{k}_\perp)^2-1\right]\delta(\omega-\eta_y(\mathbf{k}_\perp))& \mbox{if } \left\lvert\tilde{\eta}_z(\mathbf{k}_\perp)\right\rvert<1 \\
	 0 & \mbox{otherwise. }
	\end{cases}
\end{equation}
\normalsize

Since $\epsilon_0=0$, the Weyl points are located at energy $\omega=0$.  There, all the surface spectral weight is concentrated on a limited portion of the surface Brillouin zone: the Fermi arcs. The analytical expression for the location of these open equi-energy lines is then given by:
\small
\begin{align}
	\label{eqn:arcs}
	\cos{\varphi}\left(\cos{k_y}-\cos{b_y}\right)-&\sin{\varphi}\left(\cos{k_z}-\cos{b_z}\right)=0\,, \\ 
	\lvert1+\sin{\varphi}\left(\cos{k_y}-\cos{b_y}\right)&+\cos{\varphi}\left(\cos{k_z}-\cos{b_z}\right)\rvert<1\,.
\end{align} 
\normalsize

For $\varphi\in [\pi\,,5/4 \pi)$ Weyl points of opposite chirality $\chi$ with the same $k_y$, hence same $\xi$, are connected by Fermi arcs. On the other hand, for $\varphi\in (5/4 \pi\,, 3/2\pi]$, Weyl points with the same $k_z$ and opposite $\xi$ are linked on the surface. Only for $\varphi=\pi$ and $\varphi=3/2\pi$ the arcs are straight. Fig.~\ref{fig:fig1} shows the agreement between the analytical results of Eq.~\eqref{eqn:arcs} and the numerical spectral density of a finite slab on the surface Brillouin zone.

Intuitively, Weyl orbits can occur only if bulk channels with opposite group velocities are linked by Fermi arcs. This happens for $\varphi\in (5/4 \pi\,, 3/2\pi]$, when Weyl points with opposite $k_y$, and therefore opposite $\xi$, are connected by the surface states. 

A simple picture is available. Fermi arcs always link Weyl points of opposite $\chi$. Nevertheless, whether Weyl nodes with the same $\xi$ are linked is purely a matter of energetics. Whenever opposite $\chi$ and opposite $\xi$ are linked, the action of an axial field for such pair is analogous to a regular magnetic field in the low-energy sector. Time-reversal symmetry is preserved as the other pair experiences a field of opposite sign. Therefore, if one neglects scattering between different Weyl nodes, the low-energy phenomenology resembles one of a single pair of Weyl nodes in the presence of an external magnetic field and Weyl orbits should be in principle possible.

\section{Wave-packet dynamics}
\label{sec:wp}
\subsection{Exact evolution of wave-packets in a finite sample}
To provide evidence for Weyl orbits in a time-reversal invariant system without external magnetic fields, we first numerically study the evolution of wave-packets in a finite sample \cite{Yao:2017,Roy:2015}. At the initial time $t=0$, we prepare a Gaussian wave-packet localized at momentum $\mathbf{k}_0$ and centered around $\mathbf{r}_0$:
\begin{equation}
	\ket{\psi(t=0)}=\frac{1}{\mathcal{N}}\sum_j e^{-\frac{\lvert \mathbf{r}_j-\mathbf{r}_{0}\rvert^2}{2\sigma^2}}e^{i\mathbf{k_0}\cdot\mathbf{r}_j}\ket{j}\,,
\end{equation}
where $\mathcal{N}$ is a normalization factor, $\mathbf{r}_j=(x_j\,,y_j\,,z_j)$ is the coordinate vector of site $j$, $\ket{j}$ is the single-particle state fully localized at site $j$ and $\sigma^2$ is the real space variance of the initial wave-packet. The latter can differ among the various spatial components. Note that an additional orbital index has been suppressed, as the two inequivalent orbital degrees of freedom are treated on equal footings. The factor $\text{exp}(i\mathbf{k_0}\cdot\mathbf{r}_j)$ localizes the wave-packet in momentum space around $\mathbf{k_0}$. The variance in momentum space $\sigma_k^2$ is obtained from the real space one: $\sigma_k^2=1/4\sigma^2$, i.e., the more spread the wave-packet is in real space, the better localized it is in momentum space. It is important to localize the wave-packet in a single-band chiral channel induced by the axial magnetic field. To this end, we launch the wave-packet for a time $t^*$ with an extra phase factor $\text{exp}(-i\epsilon_0 t)$, where $\epsilon_0$ is the energy of the unperturbed Weyl points.
\begin{figure}[t]
\includegraphics[]{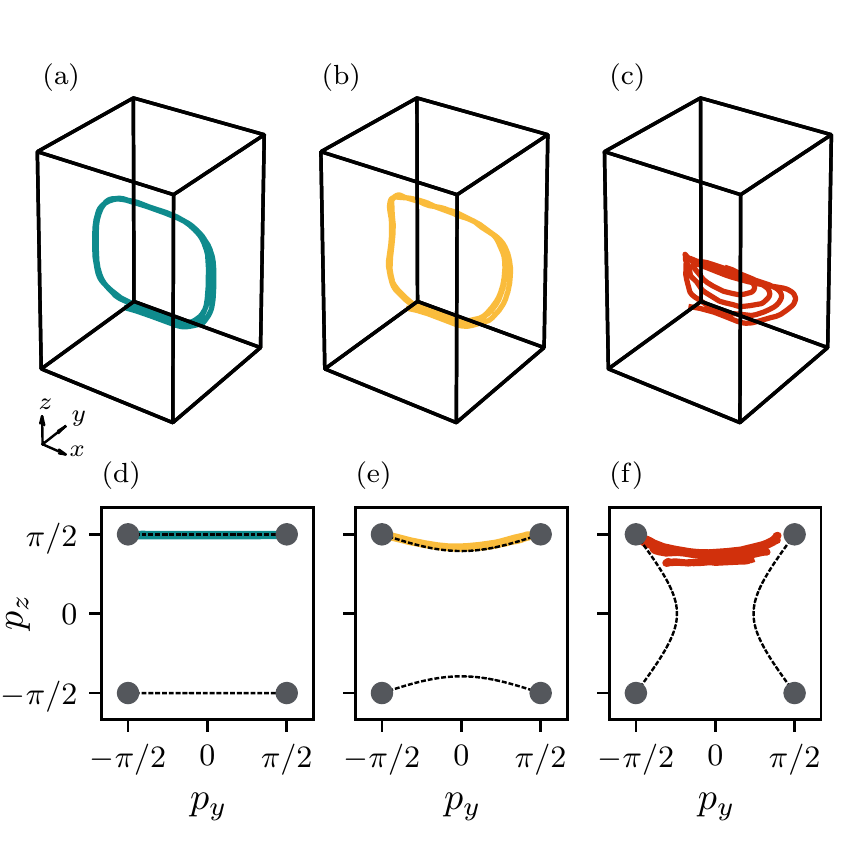}
\caption{\label{fig:fig4} In the first row: real space trajectory of a wave-packet in the presence of an axial gauge field and different values of $\varphi$. In (a) $\varphi=1.5\pi$ (green), in (b) $\varphi=1.4\pi$ (yellow) and in (c) $\varphi=1.2\pi$ (red). Weyl orbits can be clearly seen in (a) and (b). In the second row the trajectories of the kinetic momentum components $p_y$ and $p_z$ during the wave-packet evolution. The values of $\varphi$ are the same as in the first row. The black dashed lines show the analytical form of the Fermi arcs as given by Eq.~\eqref{eqn:arcs}.}
\end{figure}
During the launching procedure and the subsequent free propagation, the wave-packet is evolved according to
\begin{equation}
	\ket{\psi(t+\Delta t)}=e^{-iH\Delta t}\ket{\psi(t)}\,,
\end{equation}
\begin{figure*}[]
\includegraphics[]{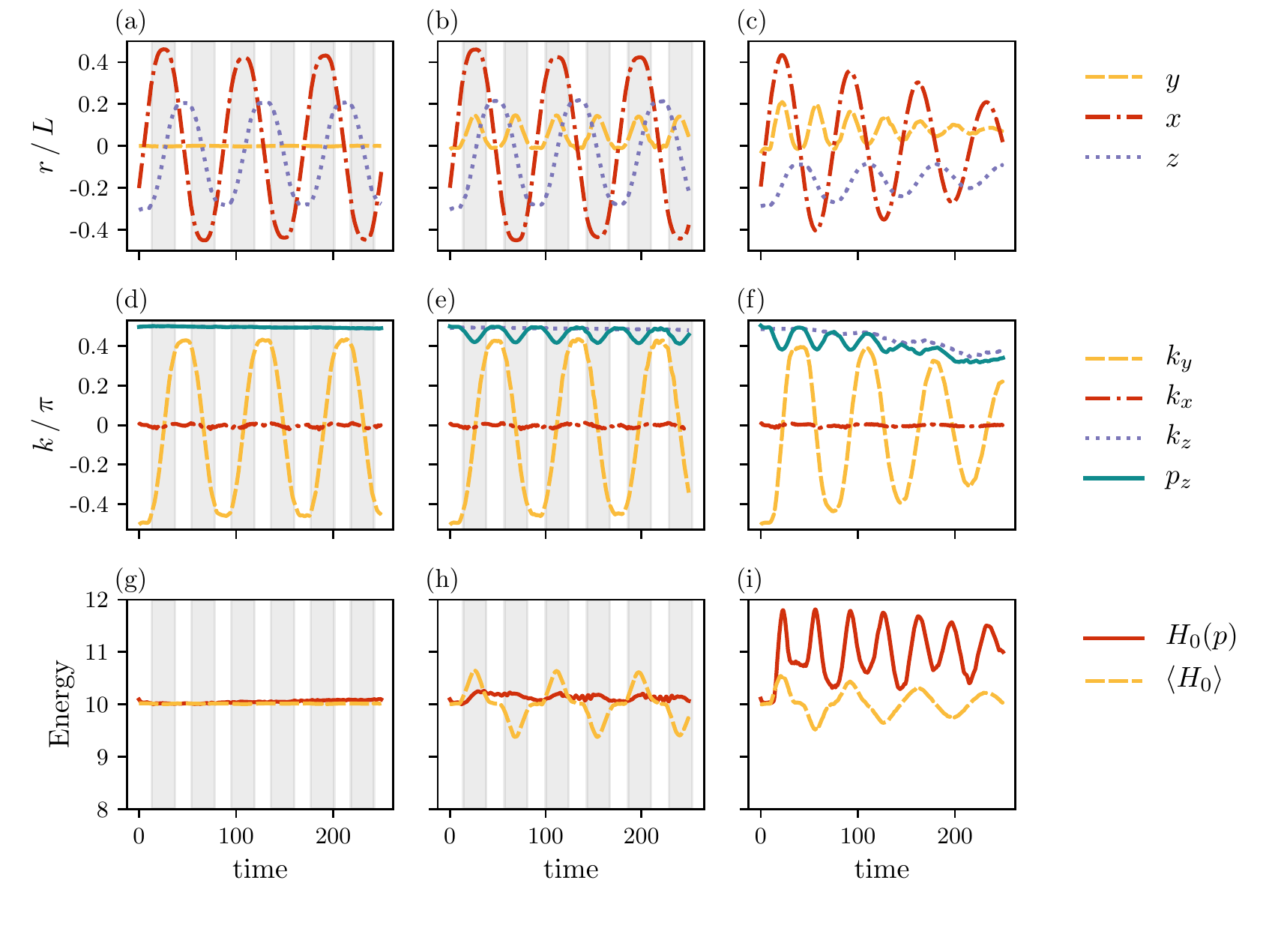}
\caption{\label{fig:fig3} Evolution of wave-packets in a finite sample $L_x=70$, $L_y=30$ and $L_z=110$ with axial field $B=\pi/60$ and different values of $\varphi$. The first column has $\varphi=1.5\pi$, the second one $\varphi=1.4\pi$ and for the rightmost $\varphi=1.2\pi$. The first rows present the real space motion of the wave-packet center of mass. The second row shows the center of mass motion in momentum space. Both the kinetic and canonical $z$ component of momentum are shown. Finally, the third row shows how the energy of the wave-packet evolves during the evolution. In particular, we show the expectation value of the unperturbed Hamiltonian as computed from the wave-packet and the eigenvalue of a slab Hamiltonian evaluated at the kinetic momentum. The grey areas in the first two columns indicate the regions where the wave-packet is mainly localized at the sample's surface.}
\end{figure*}
where $H$ is the real-space version of Hamiltonian~\eqref{eqn:model} for a finite sample of size $L_x\times L_y \times L_z$ with the space-dependent perturbations~\eqref{eqn:axial1}--\eqref{eqn:axial2}. Sparse matrices and direct numerical integration allow to perform this task efficiently. The linear system
\begin{equation}
	\left(1+\frac{iH\Delta t}{2}\right)\ket{\psi(t+\Delta t)}=\left(1-\frac{iH\Delta t}{2}\right)\ket{\psi(t)}\,
\end{equation}
is solved with biconjugate gradient iteration.

To study the evolution, we evaluate the expectation value and variance of various operators $\hat{\mathcal{O}}$: 
\begin{align}
	\expval{\hat{\mathcal{O}}(t)}&=\expval{\hat{\mathcal{O}}}{\psi(t)}\,,\\
	\Delta\hat{\mathcal{O}}^2(t)&=\expval{\hat{\mathcal{O}}^2(t)}-\expval{\hat{\mathcal{O}}(t)}^2\,.
\end{align}
In particular, we monitor the Hamiltonian $H$, the unperturbed Hamiltonian without axial magnetic field $H_0$, the wave-packet's center of mass position in real space $\mathbf{r}$ and in reciprocal space $\mathbf{k}$. Note that the expectation value of $\mathbf{k}$ is obtained from the Fourier transformed wave-packet: 
\begin{equation}
	|\tilde{\psi}\rangle=\frac{1}{\mathcal{N}}\int\text{d}\mathbf{r}\, e^{-i\mathbf{k}\cdot\mathbf{r}}\ket{\psi}\,.
\end{equation}

It is important to stress that the momentum obtained via Fourier transform is the {\em canonical momentum}. This corresponds to the conjugate operator of the position operator: $\mathbf{k}=-i\nabla$, which is however not gauge-invariant. From the expectation value of the canonical momentum operator $\mathbf{k}$ and the position operator $\mathbf{r}$, we obtain the expectation value of the gauge-invariant {\em kinetic momentum} $\mathbf{p}=m\mathbf{v}$:
\begin{equation}
	\label{eqn:canonical}
	\expval{\mathbf{p}}=\expval{\mathbf{k}}-\mathbf{A}_5(\expval{\mathbf{r}})\,.
\end{equation}
In particular, with our choice of gauge $\mathbf{A}_5=B_5y\hat{\mathbf{z}}$, we get:
\begin{align}
	\label{eqn:canonical}
	\expval{p_x}&=\expval{k_x}\,,\\
	\expval{p_y}&=\expval{k_y}\,,\\
	\expval{p_z}&=\expval{k_z}-B_5\left(\expval{y}-\frac{L_y}{2}\right)\,.
\end{align}

\subsection{Low-energy semiclassical interpretation}
In the following, we study the results of the simulations for a sample $L_x=70$, $L_y=30$ and $L_z=110$ and $B_5=\pi/60$ with $\tau=3$ and $\epsilon_0=10$. We launch the wave-packet at $\mathbf{r}_0=(0\,,15\,,40)$, centered around momentum $\mathbf{k}_0=(0\,,-\pi/2\,,\pi/2)$ and energy $\epsilon_0=10$, with variance $\sigma=(3\,,3\,,7)$. In Fig.~\ref{fig:fig3}, we show the wave-packet evolution for three different values of $\varphi$: $1.5\pi$ in the first column, $1.4\pi$ in the second column and $1.2\pi$ in the rightmost column. Intuitively, we would expect to observe Weyl orbits in the first two cases where the Fermi arcs connect bulk channels with opposite group velocities, i.e. Weyl points with opposite $\xi$ and $\chi$.

The real space motion, shown in the first row of Fig.~\ref{fig:fig3}, is compatible with Weyl orbits for $\varphi=1.5\pi$ and $\varphi=1.4\pi$. The wave-packet propagates through the bulk until it reaches the surface where it slides along the arc and gets to the channel dispersing in the opposite direction. Then, it gets reabsorbed into the bulk until it reaches the other surface and the process repeats itself (see also Fig.~\ref{fig:fig4}(a)--(b)). When $\varphi=1.2\pi$, the wave-packet reaches the surface after dispersing through the bulk along the chiral Landau level but does not seem to perform any sliding there. After a few periods, the oscillatory behavior damps out. This suggests that Weyl orbits do not take place after the re-linking of Weyl points with the same $\xi$ index, occurring at $\varphi=1.25\pi$.

The momentum space evolution deserves careful analysis and is shown in the second row of Fig.~\ref{fig:fig3}. In the presence of a magnetic vector potential, one should consider the kinetic momentum $\mathbf{p}$. Indeed, the standard semiclassical equations in the presence of non-trivial Berry curvature $\Omega$ are formulated in terms of $\mathbf{p}$ \cite{Xiao:2010,Roy:2018}:
\begin{align}
		\dot{\mathbf{r}}&=\nabla_\mathbf{p}\epsilon_\mathbf{p}-\Omega_\mathbf{pr}\cdot\dot{\mathbf{r}}-\Omega_\mathbf{pp}\cdot\dot{\mathbf{p}}\,,\\
		\dot{\mathbf{p}}&=-\nabla_\mathbf{r}\epsilon_\mathbf{p}+\Omega_\mathbf{rr}\cdot\dot{\mathbf{r}}+\Omega_\mathbf{rp}\cdot\dot{\mathbf{p}}\,.
\end{align}

Nonetheless, the axial field in condensed matter systems resembles a gauge field exclusively in the low-energy theory around the Weyl nodes. Its physical origin, e.g., inhomogeneous uniaxial strain or magnetization, and effects away from the nodal points are different from the ones of an external magnetic field. One could then arbitrarily choose to consider the kinetic momentum or the canonical one. In the first case, the one considered in this study, one focuses on the low-energy description in terms of semiclassical equations and effective gauge fields. In the second case, a microscopic description that relies on the physical implementation of the axial field explains the same phenomena. Both interpretations are consistent. 

To clarify the above discussion, we can analyze the second row of Fig.~\ref{fig:fig3}. For $\varphi=1.5\pi$ the Fermi arcs are straight. This corresponds to a surface motion only along $\hat{\mathbf{z}}$. Hence, the canonical and kinetic momenta coincide since $\expval{y}=0$ (see Eq.~\eqref{eqn:canonical}). A different situation arises when $\varphi=1.3\pi$, cf. Fig.~\ref{fig:fig3}(e). Here, the real space surface motion is both along $\hat{\mathbf{z}}$ and $\hat{\mathbf{y}}$. Therefore, canonical and kinetic momenta cease to be the same. The gauge potential explicitly breaks translational symmetry in $\hat{\mathbf{y}}$ direction. Along $\hat{\mathbf{z}}$, on the other hand, the finite sample is sufficiently long to ensure approximate conservation of momentum and hence $\dot{k}_z=0$, cf. Fig.~\ref{fig:fig3}(e). The kinetic momentum $p_z$ of Eq.~\eqref{eqn:canonical}, instead, varies as a consequence of the real space surface motion along $\hat{\mathbf{y}}$. Fig.~\ref{fig:fig4}(e) shows that the kinetic momentum describes the trajectory of the Fermi arc as given by Eq.~\eqref{eqn:arcs}. The situation is more complicated for $\varphi=1.2\pi$. There, the Fermi arcs connect Weyl points with the same index $\xi$. Hence, they link points that experience gauge fields of opposite signs in their low-energy sectors. The concept of kinetic momentum is therefore ill-defined as it is not possible to define a unique field along the whole trajectory. As can be seen in Fig.~\ref{fig:fig4}(f), the wave-packet does not follow the Fermi arc even when the kinetic rather than the canonical momentum is considered. 

Finally, the evolution of the energy expectation value is shown in the third row of Fig.~\ref{fig:fig3}. This provides clarification of the puzzling results in terms of canonical vs. kinetic momenta. The Hamiltonian in the presence of the axial gauge field can be written as:
\begin{equation}
	H=H_0+H_1\,,
\end{equation}
where $H_0$ is the original unperturbed Hamiltonian of Eq.~\eqref{eqn:model}, specialized to a finite sample, and $H_1$ introduces the axial field perturbation, cf. Eq.~\eqref{eqn:axial1}--\eqref{eqn:axial2}. The total energy is conserved during evolution. 

It is clear that a change in $y$ causes a change in the energy associated with $H_1$ since the latter depends on the gauge potential, cf. Eq.~\eqref{eqn:axial1}--\eqref{eqn:axial2}. Indeed, a motion along $y$ occurs at the sample's surface for $\varphi \neq 3/2\pi$. In turn, this implies that the expectation value of $H_0$, the unperturbed Hamiltonian of Eq.~\eqref{eqn:model}, has to change along the Fermi arc in order to keep the total energy constant. Fig.~\ref{fig:fig3}(h) confirms this observation. If one simply computes $\expval{\hat{H}_0}{\psi(t)}$, the wave-packet does not seem to follow equi-energy contours of $H_0$. This is at odds with the expectation that the addition of gauge fields does not change the energy of the wave-packet. However, this reasoning requires to consider the kinetic momentum rather than the canonical one. Only in this case, one can interpret the result in terms of semiclassical equations of motion and sliding along equienergy lines of the unperturbed Hamiltonian. 

When $\varphi\in (5/4\pi,3/2\pi]$, Weyl points with opposite $\xi$ are linked on the surface. This allows introducing a well-defined kinetic momentum since it is possible to consider a global gauge field along the whole orbit trajectory. We can then evaluate the unperturbed Hamiltonian of a finite system along $\hat{\mathbf{x}}$ and fully periodic along $\hat{\mathbf{y}}$ and $\hat{\mathbf{z}}$ at the kinetic momenta $H_0\left(k_y\,,k_z-B_5(y-L_y/2)\right)$. In this case, the energy does not change along the trajectory and the interpretation is consistent with the semiclassical picture of a wave-packet sliding along an equi-energy contour. This picture is completely analogous to one of Weyl orbits in an external magnetic field, cf. Fig.~\ref{fig:fig3}(h) and Fig.~\ref{fig:fig2}(a). 

An interpretation of the trajectories in terms of a unique well-defined gauge field is no longer possible for $\varphi\in (\pi,5/4\pi]$, i.e., when Fermi arcs connect Weyl nodes with the same $\xi$. Focusing on the low-energy theory, it appears that one of the Weyl nodes experiences a field $+\mathbf{B}_5$, while the other an opposite one $-\mathbf{B}_5$ along the trajectory. A unique kinetic momentum is now ill-defined. This is proved also by the results shown in Fig.~\ref{fig:fig3}(i), where one sees that the unperturbed Hamiltonian evaluated at the kinetic momentum does not yield a constant expectation value. The interpretation of the wave-packet motion in terms of sliding on equi-energy lines driven by a gauge field no longer holds and Weyl orbits do not take place. 

The wave-packet evolution supports the idea that Weyl orbits are possible in the exclusive presence of an axial gauge field and Fermi arcs that connect counter-propagating chiral bulk channels. The same bulk field with the ``wrong'' connectivity of the Fermi arcs, however, does not lead to Weyl orbits. 

\begin{figure*}[t]
\includegraphics[]{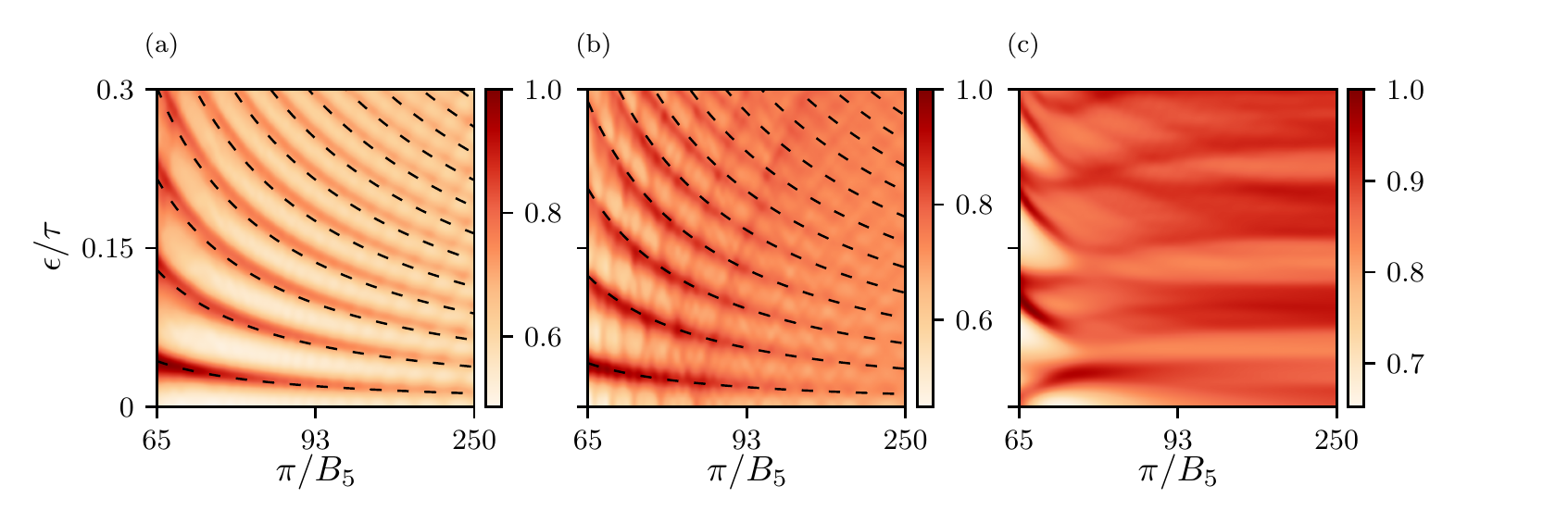}
\caption{\label{fig:fig5} Surface spectral function $S(\omega_n,\mathbf{k}_z)$ integrated over $\mathbf{k}_z$ as a function of $1/B_5$ for different values of $\varphi$. In (a) $\varphi=1.5\pi$, in (b) $\varphi=1.4\pi$ and in (c) $\varphi=1.2\pi$. The slab is characterized by $L_x=11$, $L_y=50$, $\tau=2$ and $\epsilon_0=0$. The field values vary between $\pi/250$ and $\pi/65$.
The discrete energy levels in (a) and (b) arise from the hybridization of surface Fermi arcs and bulk chiral channels confirming the presence of Weyl orbits. In (a) the black dashed line indicates the prediction of the Bohr-Sommerfeld energy quantization of Eq.~\eqref{eqn:semiQuant} with $\lambda=0.5$ and no free parameters.}
\end{figure*}

\section{Density of state oscillations}
\label{sec:dos}
The presence of Weyl orbits can be established also from the effective surface Green's functions introduced in Sec.~\ref{ssec:eff}. In gapless systems in an external magnetic field, electrons perform closed trajectories and the density of states displays oscillations as a function of the externally applied field. The oscillation frequency is inversely proportional to the area enclosed by the Fermi surface and associated to the emergence of discrete energy levels. Semiclassically, these energy levels $\epsilon_n$ satisfy the condition: $\epsilon_n T \approx 2\pi(n+\lambda)$ with $n\in\mathbb{Z}$, $T$ the period of the closed trajectory and $\lambda$ a system-dependent shift associated to possible non-trivial Berry curvature encircled by the orbit.

 The bulk Fermi surface for optimally doped Weyl semimetals is composed of isolated points. However, in the presence of field-induced Weyl orbits, the Fermi surface encloses a non-zero area. The external magnetic field $\mathbf{B}=B\hat{\mathbf{x}}$ drives electrons through the bulk along the field-induced chiral channels for a time $t=L_x/v_F$, where $L_x$ is the sample length along the field direction. Once they reach the surface, they slide along Fermi arc pushed by the Lorentz force for a time $t=k_0/ev_FB$, where $k_0$ is the length of the Fermi arc. They then get reabsorbed in the bulk and repeat the same process on the opposite surface. This allows to compute the period of the orbit $T$ and the energy discretization from the semiclassical Bohr-Sommerfeld quantization \cite{Potter:2014}:
\begin{equation}
  \label{eqn:semiQuant}
  \epsilon_n=\frac{\pi v_F}{k_0l_B^2+L_x}(n+\lambda)\,,
  \end{equation}
where the magnetic length is defined as $l_B^2=1/eB$.
It is important to stress that this semiclassical formula is obtained in the presence of an external magnetic field that generates a Lorentz force. Moreover, it assumes a constant Fermi velocity along the whole trajectory.

Effective surface Green's functions allow to investigate the energy quantization beyond the semiclassical regime and in an unbiased way. An additional axial magnetic field can be readily added to the formalism of Sec.~\ref{ssec:eff}. With the choice of Sec.~\ref{ssec:axial4} for the pseudo-magnetic potential, the translation symmetry along $\hat{\mathbf{y}}$ is explicitly broken. Hence, $k_y$ is not a good quantum number and we work in real space rather than momentum space along the $\hat{\mathbf{y}}$ direction. Therefore, the single-layer matrix $H_\perp(\mathbf{k}_\perp)$ [Eq.~\eqref{eqn:Hperp}] and the matrix that couples different layers $R$ [Eq.~\eqref{eqn:rmatrix}] become $2L_y\times 2L_y$ matrices. All the other matrices are adapted subsequently. We consider a finite slab along the field direction $\hat{\mathbf{x}}$. Compared to the semi-infinite sample of Sec.~\ref{ssec:arcsExact}, an analytical solution is no longer possible. Nevertheless, the effective surface spectral density $S_{\text{eff}}(\omega,k_z)$ can be obtained by numerical means. 

The integrated effective surface spectral density $S_{\text{eff}}(\omega)=\int_{-\pi}^{\pi} S_{\text{eff}}(\omega,k_z) \text{d}k_z$ shows maxima at the energies of the system eigenmodes. The presence of discrete energy levels is clearly seen in Fig.~\ref{fig:fig5} for values $\varphi=1.5\pi$ and $\varphi=1.4\pi$. On the other hand, for $\varphi=1.2\pi$, the spectrum does not show a discretization compatible with Eq.~\eqref{eqn:semiQuant}. Significantly, in this case, there is little dependence on the magnitude of $\mathbf{B}_5$ and the observed discretization is due to finite-size quantization. 

These observations alone suffice to establish the presence of Weyl orbits-like phenomena in time-reversal invariant setting under the application of an axial field, provided that Weyl nodes with counter-propagating unidirectional bulk channels are linked by the surface Fermi arcs. To consolidate the interpretation of axial-field-induced Weyl orbits, we overlay to Fig.~\ref{fig:fig5}(a)--(b) the semiclassical-theory prediction of Eq.~\eqref{eqn:semiQuant}. Note that the overlay is performed with no free parameters, using the arc length computed from Eq.~\eqref{eqn:arcs} and $\lambda=0.5$. The  small velocity correction along $\hat{\mathbf{z}}$ induced by the axial field has also been taken into account, cf. Eq.~\eqref{eqn:lowOur}. 

The agreement between our numerical results and the semiclassical prediction highlights two important features. First, the value $\lambda=0.5$ hints to a bulk-boundary oscillation that involves Weyl nodes with opposite chirality such that the total accumulated Berry phase along the orbit is zero \cite{Wang:2016b,Borchmann:2017}. Second, the unbiased numerical results are compared with a formula based on a real magnetic field and assuming sliding on the arc induced by Lorentz force. This confirms that the axial field can be interpreted as a magnetic-like gauge field in the low-energy theory also on the surface, at least when Fermi arcs connect Weyl points with opposite $\xi$ index. Indeed, only in this case a unique field value can be assigned to the whole orbit. 

\section{Conclusion}
\label{sec:conclude}
We presented a mechanism to realize non-local Weyl orbits while preserving time-reversal symmetry, i.e., without an external magnetic field. We showed that non-local bulk-boundary oscillations can be induced by a pseudo-magnetic field that couples with different signs to Weyl points of different chirality. This allows for the observation of a 3D Hall effect in a completely time-reversal symmetric system and parallels the relationship between the 2D quantum Hall effect and its quantum spin Hall effect counterpart. Moreover, the Weyl orbits addressed in this work strongly depend on the Fermi arc connectivity on the sample surface. This is determined by energetic arguments rather than topology and can be tuned via surface potentials \cite{Morali1286}. Therefore, our proposal realizes a surface tunable switch for non-local conveyor belt motion of electrons in a Weyl semimetal \cite{Baum:2015}. Albeit we focused on a concrete tight-binding model to illustrate the key ideas of our work, the obtained results are generic and apply to any time-reversal symmetric Weyl systems with an arbitrary number of Weyl nodes.

Our theoretical predictions can be readily tested in different experimental platforms. Engineered platforms such as cold atoms \cite{Dubcek:2015,PhysRevA.94.013606}, electric circuits \cite{Lee:2018a,Lu:2019} or photonic \cite{Lu:2015,Noh:2017} and acoustic \cite{Li:2017,Ge:2018} crystals are particularly suitable. Indeed, the axial field does not require to break time-reversal symmetry nor does one have to deal with charged particles. The former would necessitate an active control while the latter constrains to work with electrically charged excitations and rule out classical metamaterials. Our ideas could also be implemented in electronic materials \cite{Lv:2015,Xu:2015a,Yang:2015}, where the axial field can be induced by different means, e.g. inhomogeneous strain and magnetization \cite{Ilan:2019,Grushin:2016,Destraz:2020}.

Towards an implementation in electronic materials, an important future direction is the extension of our results to Dirac semimetals \cite{Potter:2014}. Experimental evidence of Weyl orbits in \ch{Cd3As2} under the application of an external magnetic field has been put forward \cite{Moll:2016,Zhang:2017,Uchida:2017,Zhang:2018} and  has attracted controversies \cite{PhysRevB.99.201401}. Strain physics has been recently studied in \ch{Cd3As2} \cite{PhysRevMaterials.3.064204} with promising results. This might lead to the implementation of axial fields in Dirac semimetals and the study of the peculiar Weyl orbits of this work. 

\begin{acknowledgments}
	VP thanks Adolfo Grushin, Wei Chen and Vardan Kaladzhyan for insightful discussions. V.P., T.D., A.V. and S.D.H. are grateful for the financial support from the Swiss National Science Foundation, the NCCR QSIT and the European Research Council under the Grant Agreement No. 771503 (TopMechMat). R.I. is supported by the Israel Science Foundation (ISF grant No. 1790/18).
\end{acknowledgments}

\bibliography{ref}
\bibliographystyle{apsrev4-1}

\clearpage

\end{document}